\documentstyle[amssymb,prl, twocolumn, aps]{revtex}

\input epsf
\begin{document}

\smallskip 
\twocolumn[\hsize\textwidth\columnwidth\hsize\csname@twocolumnfalse\endcsname

\title{Charge noise analysis of an AlGaAs/GaAs quantum dot using transmission-type
radio-frequency single-electron transistor technique}
\author{Toshimasa Fujisawa$^{\thanks{Author to whom correspondence should be addressed; electronic mail:
fujisawa@will.brl.ntt.co.jp.}}$ and Yoshiro Hirayama$^{\thanks{Also at CREST, 4-1-8, Honmachi, Kawaguchi, Saitama, 331-0012, Japan}}$}
\address{NTT Basic Research Laboratories\\
3-1, Morinosato-Wakamiya, Atsugi, Kanagawa, 243-0198, Japan}
\date{Recieved 17 January 2000;}
\maketitle

\begin{abstract}
Radio-frequency (rf)- operated single-electron transistors (SETs) are
high-sensitivity, fast-response electrometers, which are valuable for
developing new insights into single-charge dynamics. We investigate
high-frequency (up to 1 MHz) charge noise in an AlGaAs/GaAs quantum dot
using a transmission-type rf-SET technique. The electron capture and
emission kinetics on a trap in the vicinity of the quantum dot are dominated
by a Poisson process. The maximum bandwidth for measuring single trapping
events is about 1 MHz, which is the same as that required for observing
single-electron tunneling oscillations in a measurable current ($\sim $0.1
pA).
\end{abstract}


] \narrowtext

\newpage

A single electron transistor (SET) is a high-sensitivity electrometer that
measures a small fraction of the elementary charge, $e$, on a small island 
\cite{SET}. If the island, or quantum dot atom, shows well-defined
eigenstates in a zero-dimensional confinement potential, quantum mechanical
effects found in atomic physics can be reproduced with tunable parameters 
\cite{QDotReview}. The ability to control single-particle quantum states
will be helpful for developing quantum logic gates \cite
{Loss,Oosterkamp,Nakamura}. So the dynamical behavior of charge states need
to be studied for those interested. Most of the experiments have been done
by taking the dc response from an ac modulation. However, direct measurement
of the ac response of a single-particle state has not been done due to
technical difficulties. Recently, a radio-frequency (rf) operated SET
technique for following the fast response of the charge has been proposed
and demonstrated by R. J. Schoelkopf {\it et al.} \cite{Schoelkopf} . A
bandwidth greater than 100 MHz would be useful for studying single-electron
dynamics such as single-electron tunneling oscillation \cite{Averin} and
coherent charge oscillation \cite{Nakamura}, as well as for various sensors 
\cite{Schoelkopf-PD}.

In this letter, we describe an application of the rf-SET for detecting
individual emission and capture events of a trap in a semiconductor. A
modified rf-SET technique that measures the transmission of rf signals
through a resonator, is also used for a quantum dot fabricated in an
AlGaAs/GaAs heterostructure \cite{FujisawaTrfSET}. The charge noise of the
quantum dot is studied both for low-frequency $1/f$ noise and for random
telegraph signals (RTSs) originating from a trap near the dot. For a
specific RTS, the statistics of the electron capture and emission are given
by the Poisson process.

The transmission-type rf-SET technique is shown schematically in Fig. 1(a).
The SET, or quantum dot, is fabricated in an AlGaAs/GaAs modulation doped
heterostructure using focused ion beam implantation and patterning of fine
Schottky gates \cite{FujisawaPAT}. The two gate voltages, $V_{L}$ and $V_{R}$%
, control the two tunneling barriers independently, and effectively lift the
electrostatic potential of the dot. The SET is placed in an LC resonator
(two inductors of 2$L$ and one capacitor of $C$). Other lumped elements
allow measurement of the dc current and rf-transmission simultaneously. When
the rf carrier signal $V_{i}e^{i\omega t}$ is supplied at the resonant
frequency, $\omega =\frac{1}{\sqrt{LC}}$, to the resonator, a transmitted
signal $V_{t}e^{i\omega t}$ appears at the other end of the resonator. The
transmitted signal is sensitive to the resistance, $R(q)$, of the SET \cite
{impedance}, or the charge on the island, $q$, i.e., $%
V_{t}/V_{i}=-(1-4Q^{2}Z_{0}/R(q))$ for $R(q)\gg 4Q^{2}Z_{0}$ and $Q\gg 1$,
where $Z_{0}$ = 50 $\Omega $ is the cable impedance and $Q$ is the quality
factor of the resonator. The rf excitation voltage across the SET, $v_{ex}$,
is given by $v_{ex}=QV_{i}$.

Compared with the reflection-type rf-SET originally demonstrated in Ref. \ref
{Schoelkopf}, the transmission-type rf-SET has some advantages for
convenient measurements. Since the incident and the transmitted signals are
separated, a directional coupler necessary for the reflection measurement is
not required (although two inductors are necessary for the transmission
measurement). The transmission signal always shows a clear and sharp
resonance signal when the frequency is swept, while the reflection signal
does not particularly if the sample is highly resistive or Coulomb
blockaded. Once the frequency is adjusted properly, both transmitted and
reflected signals behave similarly if the sample resistance is high, $%
R(q)\gg 4Q^{2}Z_{0}$. High-sensitivity\cite{Korotkov} and high-frequency
(ideally upto $\sim \omega /Q$) operation of the rf-SET could also be
applied to the transmission-type rf-SET.

Figure 1(b) shows typical Coulomb blockade oscillations measured by dc
current (upper trace) and rf-transmission (lower trace). The traces are
taken simultaneously with a dc bias voltage of $V_{SD}$ = 0.2 mV and an rf
excitation of $v_{ex}$ $\sim $ 0.1 mV (rms for all ac amplitudes). The best
charge sensitivity is obtained at the largest slope, $|dV_{t}/dV_{g}|$,
denoted by the arrow labelled by $\alpha $\ in the figure. We estimate the
charge sensitivity, i.e., the minimal detectable charge on the island, by
applying a sinusoidal modulation to the $V_{L}$ as follows. The inset of
Fig. 2 shows the frequency spectrum of transmitted signal $V_{t}$ at the
modulation frequency, $f_{mod}$ =\ 10 kHz. The signal intensity is linearly
dependent on the modulation amplitude, $V_{mod}$, as shown in Fig. 2. We
obtain a charge sensitivity of $\delta q=3.6\times 10^{-5}e/\sqrt{Hz}$ at an
excitation of $v_{ex}$ $\sim $ 0.4 mV. This is larger than the thermal noise
and the shot noise, but is dominated by the noise of the amplifier.

Figure 3(a) shows the spectra of the transmitted signal. The thick trace
measured at the best charge sensitivity ($\alpha $ in Fig. 1(a)) shows the
SET noise, while the thin trace measured at zero conductance of the SET
shows the noise of the measurement system. The SET shows $1/f$ noise below 1
kHz, similar to that reported in metal SETs \cite{Schoelkopf,Starmark}. For
frequencies above 1 kHz, the white noise is dominated by the HEMT amplifier.
Better amplifiers and a higher resonator $Q$ should reduce the system noise.

We also find a RTS in our dot. The specific trap we study has a relatively
fast switching time and is a good example for demonstrating the fast
response of the rf-SET. Figure 1(c) shows another CB oscillation with a jump
seen at about -710 mV, labelled $\beta $ in the figure, where the jump
appears at the largest $|dV_{t}/dV_{g}|$ of the CB peak for the best
sensitivity. The jump is caused by the emission and capture of an electron
from a trap in the vicinity of the quantum dot. The trap may be a defect in
the crystal or a potential hollow (maybe a quantum dot) unintentionally
created during fabrication. This specific trap is located close to the left
gate rather than the right gate (schematically shown in the left inset of
Fig. 4(a)) \cite{trap}.

The RTS noise is observed on a relatively short time scale for this
particular trap, as shown in Fig. 4(a). The emission and capture statistics
are derived from the RTS noise in the time domain. The duration time for the
trap being empty (lower signal in the figure) corresponds to the capture
rate, and the duration time for the trap being occupied (higher signal)
corresponds to the emission rate. These durations are widely distributed, as
shown in Fig. 4(b), and well characterized by an exponential dependence, $%
\exp (-t/\tau )$, with time constant $\tau $. The mean ($m$)\ and the
standard deviation ($\sigma $)\ of the durations are practically the same as 
$\tau $. These observations indicate a Poisson process, i.e., each emission
and capture process is an independent event \cite{Kirton}. The time constant
for capture, $\tau _{c}$, and emission, $\tau _{e}$, are strongly dependent
on the energy level of the trap, $E_{t}$, or on the gate voltage $V_{L}$
(see Fig. 4(c)). Simply assuming a constant attempt rate for both
transitions, $\tau _{0}^{-1}$, and a Fermi distribution, $f_{FD}(E)\equiv
(1+\exp (E/k_{B}T_{eff}))^{-1}$, in the source electrode, the capture and
emission rates should be given by $\tau _{c}^{-1}=\tau
_{0}^{-1}f_{FD}(E_{t}-E_{F})$ and $\tau _{e}^{-1}=\tau
_{0}^{-1}(1-f_{FD}(E_{t}-E_{F}))$, respectively. The measured time-constants
are reproduced with parameters $\tau _{0}=$ 12 $\mu $s and $k_{B}T_{eff}$ =
0.46 meV, as shown by the solid lines in Fig. 4(c). The effective
temperature, $T_{eff}$, is comparable to the rf excitation voltage across
the SET, $v_{ex}$ $\sim $ 0.4 mV. We find that $T_{eff}$ increases with $%
v_{ex}$.

The power spectrum of the RTS noise (Fig. 3(b)) is flat up to a few kHz and
decreases above 5 kHz. The spectrum can be fitted with a Lorentzian curve
(dashed line), $\varpropto (1+(2\pi f\overline{\tau })^{2})^{-1}$. The
fitting parameter, $\overline{\tau }$ = 16 $\mu $sec, is comparable to $\tau
_{e}\tau _{c}/(\tau _{e}+\tau _{c})$ from the time domain analysis, which
also indicates a Poisson process.

The RTS noise has been studied in sub-micrometer Si
metal-oxide-semiconductor structures \cite{Kirton}, AlGaAs/GaAs narrow
channels \cite{Liefrink}, and quantum dots \cite{Sakamoto}. A higher sample
resistance in general restricts the lower frequency range that can be
measured. Thus, most of the measurements on SETs ($R$ 
\mbox{$>$}%
26 k$\Omega $ in principle) are restricted to the $1/f$ noise region \cite
{Starmark}. Our observation of RTS in the quantum dot is consistent with
those previously reported. The maximum bandwidth $B$, where the change in
the charge, $\Delta q$, is distinguished from the noise, $\delta q$, is
given by $B\sim \frac{1}{8}(\Delta q/\delta q)^{2}$. This yields $B$ $\sim $%
\ 1 MHz for the trap we investigated. We observe a duration as short as few $%
\mu \sec $ and rise/fall time shorter than 1 $\mu \sec $ (the right inset of
Fig. 4(a)), when the bandwidth is set to 2.5 MHz. If the trap is replaced
with another quantum dot connecting another set of leads \cite{Duncan},
single electrons will tunnel through the dot with a frequency of $I/e$ (600
kHz at a driving current, $I$, of 0.1 pA). The rf-SET should resolve each
tunneling event of such a current.

In summary, we have demonstrated a transmission-type rf-SET using a quantum
dot for detecting charge kinetics of a trap in an AlGaAs/GaAs
heterostructure. Our results are encouraging for the direct measurement of
ac response from a single-electron state.

We thank D. G. Austing, R. Espiau de Lamaestre, A. Kanda, S. Komiyama, B.
Starmark for their discussion and help.


\newpage

\begin{figure}
\epsfxsize=3.2in \epsfbox{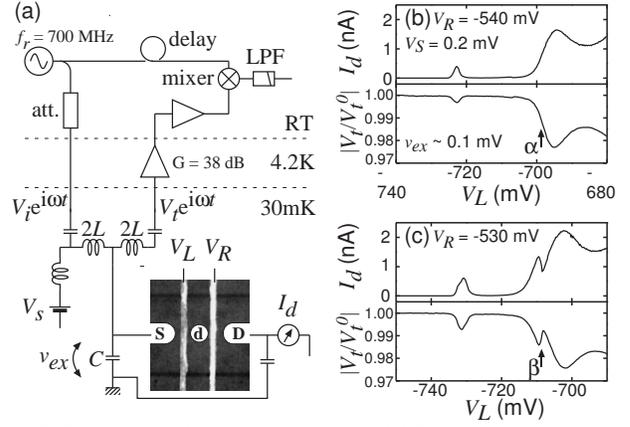}
\caption{(a) Schematic setup for rf-SET measurements. The quantum dot, d,
fabricated in an AlGaAs/GaAs heterostructure is controlled with two gate
voltages, $V_{L}$ and $V_{R}$. The charging energy of the dot is about 3
meV. The SET is placed in an LC resonator ($L$ = 100 nH, $C$ $\sim $ 0.5 pF,
resonant frequency $\omega /2\pi $ = 700 MHz, and quality factor about 4).
Signal $V_{t}$e$^{i\omega t}$ is amplified and detected with a homodyne
configuration. The sample is cooled to 30 mK and measured at zero magnetic
field. (b) Typical Coulomb blockade (CB) oscillation measured by dc current
(upper traces) and rf-transmission (lower traces). (c) CB oscillation in
another case where the SET probed a trap.}
\end{figure}

\begin{figure}
\epsfxsize=3.0in \epsfbox{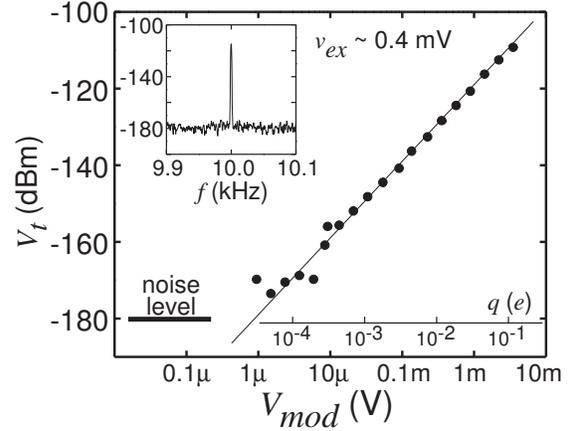}
\caption{Transmitted signal $V_{t}$ measured by applying a small modulation
signal, $V_{mod}\sin (2\pi f_{mod}t)$, to the left gate ($f_{mod}=$ 10 kHz).
Static gate voltages are $V_{L}$ = -700 mV and $V_{R}$ = -540 mV (labelled $%
\alpha $ in Fig. 1(b)). Transmitted power at $f_{mod}$ is plotted against $%
V_{mod}$. The horizontal bar indicates the noise level measured with a
resolution bandwidth of 1 Hz. The inset shows the spectrum of the
transmitted signal at $V_{mod}$ = 1.7 mV.}
\end{figure}

\bigskip

\begin{figure}
\epsfxsize=3.2in \epsfbox{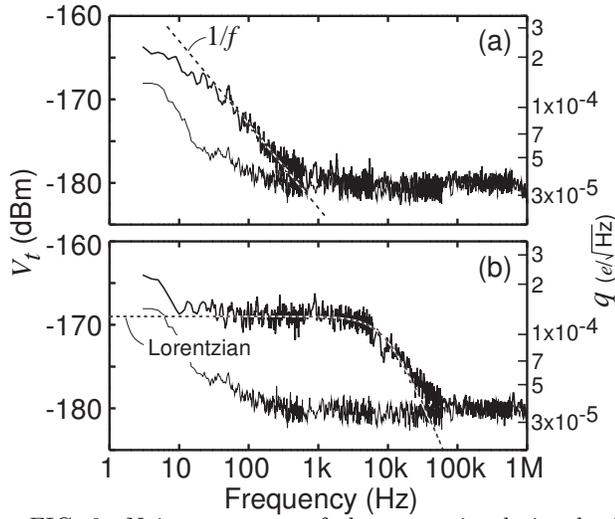}
\caption{Noise spectrum of the transmitted signal. The equivalent charge with
respect to the quantum dot is on the right. (a) Noise of the SET
electrometer. Thick trace shows the SET noise measured at $\alpha $ in Fig.
1(b), while the thin trace indicates noise from the measurement system
(taken where SET is highly resistive). The dashed line shows a $1/f$
dependence. (b) Noise when the random telegraph signal (RTS) is dominant ($%
\beta $ in Fig. 1(c)). The thin line is the system noise. The dashed curve
is a Lorentzian fitted to the RTS spectrum.}
\end{figure}

\bigskip

\begin{figure}
\epsfxsize=3.0in \epsfbox{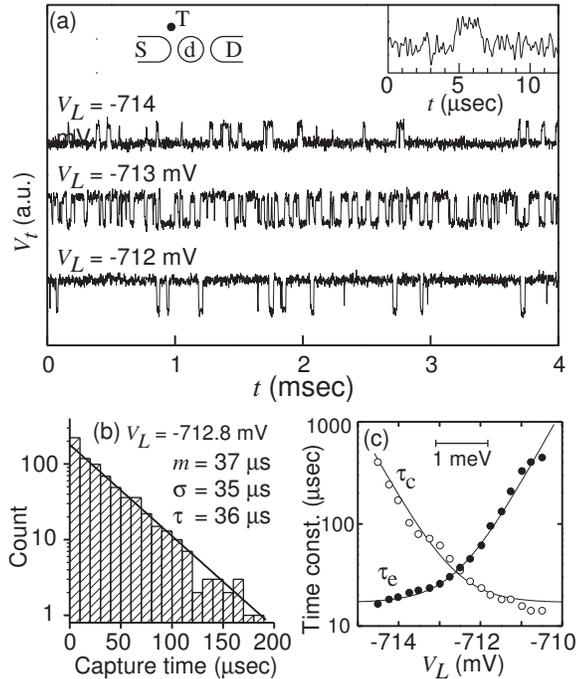}
\caption{(a) Random telegraph signal (RTS) measured at different gate
voltages. Signal $V_{t}$ is high if the trap is occupied by an electron and
low if the trap is empty. The right inset shows the shortest duration of the
occupied state. The left inset shows the location of the trap T relative to
the dot d. (b) Typical histogram of the duration of the empty state. The
solid line is exponentially fitted to the histogram. (c) Time constants $%
\tau _{c}$ for capture and $\tau _{e}$ for emission. The horizontal bar is
the energy scale of the trap state, $E_{t}$. The solid lines are fitted from
the Fermi distribution of electrons in the leads.}
\end{figure}


\begin{references}
\bibitem{impedance}  The SET consists of tunneling resistance and
capacitance, whose values can be deduced from the phase analysis of the
transmitted signal. We focus here on the region where the resistance change
of the SET is dominant.
\bibitem{trap}  The location of the trap can be estimated from the
gate-voltage dependence, as reported in Ref. \cite{Sakamoto}. The energy
state of the trap is tunable with $V_{L}$. The trap is occupied by an
electron for $V_{L}>$ -710 mV and is empty for $V_{L}<$ -710 mV. The
trapping of an electron lifts the potential of the dot by 0.3 meV, or
induces an equivalent charge of 0.1$e$ on the dot, and changes (about 20\%)
the resistance of the tunneling barriers.
\end{references}
\end{document}